\documentclass{article}

\usepackage{arxiv}

\usepackage[utf8]{inputenc} 
\usepackage[T1]{fontenc}    
\usepackage{hyperref}       
\usepackage{url}            
\usepackage{booktabs}       
\usepackage{amsfonts}       
\usepackage{nicefrac}       
\usepackage{microtype}      
\usepackage{lipsum}		
\usepackage{graphicx}
\usepackage[numbers]{natbib}
\usepackage{doi}

\title{Turing’s First Imitation Game: Design Concepts and a Human-Approximates-Machine Reading}

\author{\hspace{1mm}Sharon~Temtsin \\
	University of Canterbury \\
	Christchurch, New Zealand \\
	\texttt{sharon.temtsin@pg.canterbury.ac.nz} \\
	\And
	\hspace{1mm}Christoph Bartneck \\
	University of Canterbury\\
	Christchurch, New Zealand  \\
	\texttt{christoph.bartneck@canterbury.ac.nz} \\
}

\renewcommand{\shorttitle}{\textit{arXiv} Template}

\hypersetup{
pdftitle={A template for the arxiv style},
pdfsubject={q-bio.NC, q-bio.QM},
pdfauthor={David S.~Hippocampus, Elias D.~Striatum},
pdfkeywords={First keyword, Second keyword, More},
}

\begin{document}
\maketitle

\begin{abstract}
	This paper examines Turing's 1948 report, ``Intelligent Machinery'', as an important conceptual source for the later imitation games. Its first contribution is to identify and integrate the design concepts underlying the 1948 chess-based imitation game: the possibility that intelligent machines may make mistakes, the exclusion of irrelevant physical features, the role of the human judge, and Turing's claim that intellectual activity consists mainly of search. The paper's second contribution is to argue that restricting the human contestant to a rather poor chess player increases the role of intellectual search and makes human behaviour more comparable to machine behaviour. This interpretation presents the 1948 game as a human-approximates-machine game and suggests that the imitation game framework can be used not only to ask whether machines imitate humans, but also to examine when human intelligence becomes machine-like under specific task constraints.
\end{abstract}


For more than 75 years, the Turing Test has challenged machines to show their thinking capability \citep{1950IG, Temtsin2025}. In the game, the interrogator communicates textually with two hidden contestants and attempts to determine which is the machine and which is the human. If the machine causes the interrogator to make the wrong identification, the result supports a positive answer to the question ``Can machines think?'' \citep{1950IG}. In 1952, Turing also described a single-contestant version in which either a human or a machine is judged by a jury \citep{1952IG}. 

The 1950 paper was not Turing’s first use of the imitation-game form \citep{Copeland2000, 1948IG}. In his 1948 report ``Intelligent Machinery'', written for the National Physical Laboratory, Turing proposed to investigate ``whether it is possible for machinery to show intelligent behaviour.'' \citep[p.~107]{Turing1948_original_citation}. The report also introduced ideas that today are the basics of modern Artificial Intelligence (AI), such as discrete-state machines that simulate biological neurons, neural network training, and reinforcement learning \citep{Connectionism_TT1948_CP1996}. At the end of the report, he described the chess-based imitation game: ``It is possible to do a little experiment on these lines, even at the present stage of knowledge. It is not difficult to devise a paper machine which will play a not very bad game of chess. Now get three men as subjects for the experiment A, B, C. A and C are to be rather poor chess players, B is the operator who works the paper machine. (In order that he should be able to work it fairly fast it is advisable that he be both mathematician and chess player.) Two rooms are used with some arrangement for communicating moves, and a game is played between C and either A or the paper machine. C may find it quite difficult to tell which he is playing.
(This is a rather idealized form of an experiment I have actually done.)''\citep[p.~127]{Turing1948_original_citation}. Unfortunately, the report was unpublished during Turing's lifetime (only in 1968 \citep{Connectionism_TT1948_CP1996}).

Although some individual aspects of Turing's imitation games have been discussed separately, the 1948 report has not been systematically examined as presenting a coherent set of design concepts. One of Turing's design concepts was to allow a thinking machine to make mistakes \citep{Turing1948_original_citation, 1950IG}. A game such as chess provides a useful framework for this concept because a player may make imperfect moves and still remain a competent player, or even win the game. Similarly, a machine does not need to produce perfect behaviour in every textual reply to be evaluated as intelligent in the language-based imitation game.

 Another concept states that physical features should not contribute to intelligence evaluation  \citep{Turing1948_original_citation, 1950IG}. In 1948, Turing rejected the idea of building a complete artificial human body as too slow and impractical \citep{Turing1948_original_citation}. Instead, he proposed investigating a ``brain'' without a body, at most equipped with organs of sight, speech, and hearing \citep{Turing1948_original_citation}. Moreover, later he claimed that the imitation game makes a machine's physical disabilities  ``irrelevant'' \citep{1950IG}. Isolating the machine, and allowing textual communication only prevents the judge from relying on embodiment, voice, appearance, or other features that do not directly test the relevant intellectual behaviour.

In addition, textual communication addresses Turing's view that intelligence recognised through produced behaviour. In 1948, Turing described intelligence as composed of two factors, \textit{discipline and initiative}. Turing wrote: ``To convert a brain or machine into a universal machine is the extremest form of discipline. Without something of this kind one cannot set up proper communication.'' \citep[p.~125]{Turing1948_original_citation}. Discipline refers to the rules or structure that make communication and action possible. Following discipline as a baseline, he explained what is initiative: ``But discipline is certainly not enough in itself to \textit{produce intelligence}. That which is required in addition we call initiative.'' \citep[p.~125]{Turing1948_original_citation}.

Turing associated initiative with problems of search: ``A very typical sort of problem requiring some sort of initiative consists of those of the form `Find a number n such that . . .'.'' \citep[p.~126]{Turing1948_original_citation}. He theorised that ``intellectual activity consists mainly of various kinds of search'' \citep[p.~127]{Turing1948_original_citation}, later promoted by Newell and Simon \citep{Connectionism_TT1948_CP1996}. Turing discussed four forms of search: ``the crudest way'' \citep[p.~126]{Turing1948_original_citation} this is what we call ``the brute force'' search where the search goes through all possible options, Turing considered this method ``successful in the simplest cases'' \citep[p.~126]{Turing1948_original_citation}, genetical search; cultural search; and intellectual search, defined as ``searches carried out by brains for combinations with particular properties'' \citep[p.~127]{Turing1948_original_citation}.  

On this interpretation, intellectual search is a process through which intelligent behaviour is generated. It consists of operations carried out by a brain or machine, using rules and memory, to produce an output. If intellectual activity consists mainly of search, then intelligence is expressed in what search produces. The imitation game, therefore, does not evaluate the internal search process directly but the product of that process as expressed through textual interaction. Turing reinforced this idea in 1950 choosing ``question and answer method'' as a way ``for introducing almost any one of the fields of human endeavour'' \citep[p.~435]{1950IG}. 

Turing also described intelligence as an ``emotional rather than mathematical'' \citep[p.~108]{Turing1948_original_citation}. He returned to this point in the final section of the 1948 report, ``Intelligence as an Emotional Concept'', where he argued that intelligence is attributed when the produced behaviour resists simple explanation or prediction by the human observers: ``If we are able to explain and predict its behaviour or if there seems to be little underlying plan, we have little temptation to imagine intelligence” \citep[p.~127]{Turing1948_original_citation}. On this view, human perception of observable behaviour is part of the evaluation of intelligence, motivating the judge's role in the design of the imitation game. Yet this does not mean that any judge is equally suitable. Turing later specified that “a jury … should not be expert about machines” \citep[p.~495]{1952IG}. This restriction supports the description of the judge as ``rather poor'' in the 1948 game \citep{1948IG, Copeland2000} and as an ``average interrogator'' in interpretations of the 1950 game \citep{1950IG, Copeland2000}.

In 1948, Turing chose chess as the medium of interaction for the imitation game because moves can be communicated textually through standard notation, without requiring physical embodiment. It also requires disciplined action within a rule-governed system, including the goal of checkmate, the permitted movement of pieces, and the formal structure of legal play, while demanding intellectual search in the selection of moves. Moreover, the game places the judge in the position of an opponent, allowing them to observe and evaluate each contestant through the chess moves they produce. 

The role of the chess-based imitation game in the 1948 report remains underexplained. Some accounts suggest that Turing used the game to ``set the scene to investigate'' the report’s central question \citep[p.~1]{Watson_Shah11}, and that it should be understood as an early form of the imitation game later developed in 1950 \citep{Copeland2000, 1948IG}. However, this explanation leaves the answer to the 1948 question open, even though Turing stated that the experiment was ``actually done''\citep[p.~127]{Turing1948_original_citation}. Another interpretation holds that Turing assumed chess requires intelligence and therefore introduced a restricted computer-imitates-human game, since chess cannot represent the full range of intelligent behaviour, the game emphasises the role of the human judge in evaluating intelligence \citep{RestCIHG_Proudfoot13}. It is conceivable that Turing considered that playing chess required intelligence from humans, yet this still does not explain why Turing specifically restricted the human reference contestant to be ``a rather poor'' chess player. If the purpose were only to show that chess requires intelligent effort, then a player at any level could serve as the human reference.

A different interpretation follows from Turing's 1948 hypothesis that ``It will only be when the man is `concentrating' with a view to eliminating these stimuli or `distractions' that he approximates a machine'' \citep[p.~118]{Turing1948_original_citation}. Chess requires this kind of concentration: human players must suppress irrelevant stimuli and maintain attention on the board to select effective moves. Empirical studies show that physical and cognitive interference can disrupt chess performance \citep{chess_second_task_Dennis1989, mask_chess_effect_Smerdon22}. The 1948 game can therefore be read as a human-approximates-machine game. The human contestant does not deliberately imitate the computer; rather, the concentrated, rule-governed conditions of chess may produce machine-like behaviour. If the interrogator cannot distinguish the contestants, as Turing reported \citep{Turing1948_original_citation}, this supports the hypothesis that humans can approximate machines under such conditions.

However, approximation to machine-like behaviour does not by itself show that the behaviour is based on intellectual search. Empirical studies indicate that chess move selection relies on both memory-based recognition and search processes: weaker players depend more heavily on search, whereas stronger players rely more on recognition of familiar board configurations \citep{chess_memoryVSSearch2014, Search_in_chess2011}. Turing's restriction of the human contestant to ``a rather poor'' chess player increases the role of intellectual search in the human contestant's decision-making, making it more comparable to the search-based behaviour of the machine contestant. In contrast, the machine contestant was described only as one ``which will play a not very bad game of chess'' \citep[p.~127]{Turing1948_original_citation}. This sets a minimal competence threshold for the machine, rather than requiring a specific level of chess expertise.

By the last sentence in the 1948 report: ``C may find it quite difficult to tell which he is playing.'' \citep[p.~127]{Turing1948_original_citation} Turing provided a positive answer to the main question that ``it is possible for machinery to \textit{show} intelligent behaviour'' \citep[p.~107]{Turing1948_original_citation}. The resulting difficulty in distinguishing the contestants implies that human intellectual behaviour, under these constraints, can appear mechanical. It does not establish that the machine is intelligent or that machine-like behaviour is unintelligent; it suggests that some forms of intelligence are themselves machine-like. The 1948 chess game therefore reveals a broader function of the imitation game as a method for comparing a machine and a human: it can test not only whether machines can imitate humans, but also whether humans, under specific task conditions, can display machine-like forms of intelligence.

This interpretation of the 1948 report suggests that human and machine behaviour can become comparable under formal rules and search-based constraints. The imitation game may therefore have particular significance for computer science tasks, since coding, debugging, algorithm tracing, and formal problem solving are central to the construction and evaluation of computational systems. In these tasks, humans already employ machine-like forms of intelligence: they follow the rules of programming languages, search through possible solutions, detect errors, and produce outputs that can be tested. The imitation game can therefore help computer science examine the conditions under which human and machine intelligence converge in computational problem solving.

\bibliographystyle{unsrtnat}
\bibliography{references}  






\end{document}